\documentclass{ws-procs961x669}            

\begin{document}
\title{Microscopic Model building for Black Hole Membranes \\
from Constraints of Symmetry}

\author{Swastik Bhattacharya $^*$ and S. Shankaranarayanan $**$}

\address{Department of Physics, BITS Pilani Hyderabad;\\ Hyderabad 500078, Telangana State, India\\
$^*$E-mail: swastik@hyderabad.bits-pilani.ac.in}

\author{S. Shankaranarayanan}

\address{Department of Physics, Indian Institute of Technology Bombay, \\
Mumbai~400076, India\\
$**$ E-mail: shanki@phy.iitb.ac.in}

\begin{abstract}
Einstein equations projected on black-hole horizons give rise to the equations of motion of a viscous fluid. This suggests a way to understand the microscopic degrees of freedom on the black-hole horizon by focusing on the physics of this fluid. In this talk, we shall approach this problem by building a crude microscopic model for the Horizon-fluid(HF) corresponding to asymptotically flat black-holes in 3+1 dimensions. The symmetry requirement for our model is that it should incorporate the S1 diffeo-symmetry on the black-hole horizon. The second constraint comes from the demand that the correct value of the Coefficient of the Bulk Viscosity of the HF can be deduced from the model. Both these requirements can be satisfied by an adoption of the eight vertex Baxter model on a S2 surface. We show that the adiabatic entropy quantisation proposed by Bekenstein also follows from this model. Finally, we argue the results obtained so far suggest that a perturbed black-hole can be described by a CFT perturbed by relevant operators and discuss the physical implications. 
\end{abstract}


\bodymatter

\section{Introduction}\label{aba:sec1}
Gravity, quantum theory, and thermodynamics~\cite{2001-Wald-LRR,2003-Jacobson.Parentani-FP,2008-Carlip-Lec} are related at some fundamental level. 
The laws of black-hole mechanics are formally similar to the Laws of Thermodynamics and suggest that black-holes have entropy and temperature~\cite{1973-Bardeen.etal-CMP,1995-Jacobson-PRL,2010-Padmanabhan-RPP,2015-Padmanabhan-Entropy,1975-Hawking-CMP}.
A satisfactory explanation of black-hole thermodynamics requires a statistical mechanical origin of entropy.  It has been argued that most of the black-hole degrees of freedom (DOF) reside on the horizon, as the black-hole entropy scales as area \cite{2001-Wald-LRR,2007-Das.Shankaranarayanan-CQG}. Also, black-hole thermodynamics deals only with equilibrium states.

Gravity causes the non-stationary black-hole horizon to interact with external fields (perturbations) continually.  These interactions lead to the transfer of energy from the external fields to black-hole degrees of freedom. If we compute observables involving black-hole, we observe \emph{dissipative} effects corresponding to interaction with fields. In macroscopic black-hole physics, this can be explicitly seen by projecting the equations of motion of external fields and gravity theory on the black-hole event horizon leading to dissipative equations. In these scenarios, fluid dynamics description is useful as only average quantities resulting from the interactions at the microscopic level are observed on macroscopic scales~\cite{2019-Romatschke.Romatschke-Book}. Interestingly, it was shown black-hole horizon behaves like a viscous fluid and satisfies Damour-Navier-Stokes equation~\cite{1982-Damour-Proc,1986-Price.Thorne-PRD,1986-Thorne.etal-Membrane}. 

Since we do not have a consistent model of quantum gravity, we explore here the viability of certain ideas in the context of a toy model for the non-stationary black-hole horizon that is slightly perturbed from stationarity. The toy model we construct takes into account \emph{two different aspects of the black-hole horizon}.   First, the model incorporates near-horizon symmetries of the stationary black-hole~\cite{1998-Kaul.Majumdar-PLB,2011-Kaul.Majumdar-PRD,2001-Koga-PRD,2001-Hotta.etal-CQG,2002-Hotta-PRD,1999-Carlip-PRL,2011-Carlip-Entropy}. Second, the model accounts for the physics of transport phenomena of horizon-fluid~\cite{2016-Bhattacharya.Shankaranarayanan-PRD,2017-Cropp.etal-PRD}.

The first aspect is the constraint of symmetry on the microscopic theory. Thus, for example, we demand that the model must incorporate near-horizon symmetries. The second aspect can be viewed as the expectation that the transport phenomena exhibited by the horizon-fluid corresponding to a black-hole in the hydrodynamic limit should follow from this model.

A Conformal Field Theory (CFT) on the black-hole horizon can partly incorporate the near-horizon symmetries and is a natural candidate for the microscopic theory of stationary black-holes. We extend the analysis and show that a perturbed CFT can also incorporate this symmetry. Thus in our model, a non-stationary black-hole can be viewed as interacting with external fields. For the microscopic description, this means adding interaction terms to the stationary black-hole described by CFT. We can see this process as a perturbed black-hole relaxing to a stationary black-hole by emitting QNMs~\cite{1992-Chandrasekhar-BHBook}. In practice, we shall deform the CFT by introducing only a mass term. However, it plays a vital role by determining the infra-red cutoff for the theory and, as a consequence, also determining the value of the coefficient of Bulk Viscosity, $\zeta$. 

We focus only on the Bulk Viscosity here, which means we only consider homogeneous processes, e.g., the increase of black-hole area in a spherically symmetric space-time due to infalling spherical mass shell. Modeling such processes is comparatively easier. Stationary, non-extremal black-holes in 4-dimensional general relativity
exhibit an infinite-dimensional symmetry in the near-horizon region~\cite{1998-Kaul.Majumdar-PLB,2011-Kaul.Majumdar-PRD,2001-Koga-PRD,2001-Hotta.etal-CQG,2002-Hotta-PRD,
1999-Carlip-PRL,2011-Carlip-Entropy,2010-Barnich.Troessaert-PRL,2016-Donnay.etal-PRL}.  Thus, the near-horizon possess infinite-dimensional algebra such as $\mathcal{S}1$ diffeomorphism~\cite{1999-Carlip-PRL,2011-Carlip-Entropy} or (near) BMS~\cite{2016-Donnay.etal-PRL,1962-Bondi.etal-PRLSA,1962-Sachs-PRLSA,2017-Strominger-Arx}. 
The CFT describing a stationary black-hole can incorporate the $\mathcal{S}1$ diffeosymmetry\cite{1999-Carlip-PRL,2011-Carlip-Entropy} as it 
possesses a representation of the Virasoro algebra~
\cite{2011-Carlip-Entropy,2012-Compere-LRR}.

The perturbed CFT we choose possesses symmetries that lead to a representation of the Virasoro algebra~\cite{1987-Zamolodchikov-JETPL}. These are \emph{Integrable field theories} with an infinite number of conserved charges corresponding to an infinite number of symmetries~\cite{1989-Zamolodchikov-Proc}. The crucial point that allows us to model the black-hole horizon-fluid by such a perturbed CFT model is that one of the representations of the Virasoro algebra corresponding to the perturbed CFT is also a representation of the $\mathcal{S}1$  diffeomorphism symmetry. Throughout what follows, we use natural units setting $\hbar = c = G = k_B = 1$. 

\section{The Microscopic Model}

We consider here the {\sl Eight-vertex Baxter model}~\cite{1971-Baxter-PRL,1973-Baxter.Wu-PRL,1976-Baxter-JStatPhys,1977-Baxter-JStatPhys,1978-Baxter-JStatPhys,2016-Baxter-Book} for our purpose. We show that this model exhibits both the two aspects discussed above. The model has the following properties that form crucial ingredients for the microscopic model building of the horizon-fluid: First, it possesses {\sl lattice} Virasoro algebra corresponding to ${\mathcal S}1$ diffeomorphism symmetry~\cite{1971-Kadanoff.Wegner-PRB,1987-Itoyama.Thacker-PRL}. Second, it consists of two staggered 2D Ising lattices, and its free energy density is the same as that of the 2D Ising model. However, the two-sublattice symmetry is very different from that of the usual Ising model. Hence,  the Baxter solution's critical indices are, in general, different from that of Ising~\cite{1971-Kadanoff.Wegner-PRB}. Third, it exhibits a second-order phase transition. In the continuum limit, it is an Integrable Field Theory near the critical point and is a CFT at the critical point~\cite{1986-Thacker-Physica,1987-Itoyama.Thacker-PRL,1989-Itoyama.Thacker-NPB}.
\begin{figure}[!hbt]
    \centering
    \includegraphics[width=\linewidth]{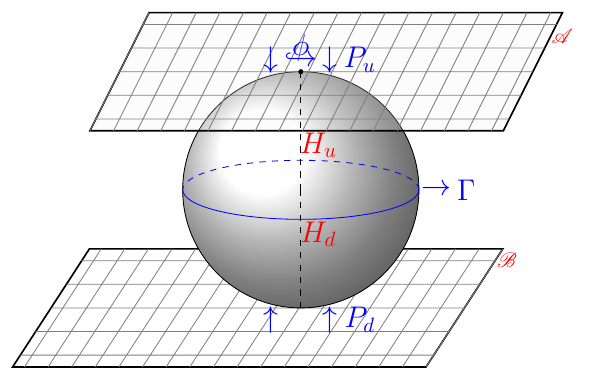}
  \caption{The projection of  the 8-vertex Baxter model from the two sub-lattices to $S^2$ surface of the horizon.} 
\label{fig:Stereographic}  
   \end{figure}
   
The Baxter model can be adapted to the cross-section of a black-hole event-horizon, a $S^2$ surface. As shown in \ref{fig:Stereographic}, this model can be adopted on the two hemispheres of the $S^2$ surface through projection from two Baxter lattices.  Let $P_u$ ($P_d$) denote the map corresponding to the projection $\mathcal{A}\rightarrow H_u$ ($\mathcal{B}\rightarrow H_d$). For the consistency of the model, we need to impose the condition $P_u^{-1}\circ\{\Gamma\} \equiv P_d^{-1}\circ\{\Gamma\}$, where, $P^{-1}$ denotes the inverse map and $\Gamma$ is equatorial plane of the $S^2$ surface. 

The above condition retains the periodic boundary condition of the Baxter model.  The projection allows relating the Euclidean boost parameter of the Baxter model~\cite{1989-Itoyama.Thacker-NPB} to the azimuthal angle in the spherical polar coordinate. We can then relate the Virasoro algebra (corresponding to the $\mathcal{S}1$ diffeomorphism) of the Euclidean boost parameter to the $\mathcal{S}1$ diffeomorphism of the azimuthal angle in the horizon-fluid model. Thus, the projection retains the model's main physical features~\cite{1989-Itoyama.Thacker-NPB} and {\it directly} incorporates a representation of the $S1$ diffeosymmetry in the model. {\it It is simply the diffeosymmetry of the azimuthal angle for the black-hole system.} 

\section{Eight-Vertex model and deformed CFT}

The Eight-vertex model has eight possible arrangements of arrows at a vertex with four distinct Boltzmann weights $a, b, c, d$. These satisfy two constraints~\cite{2016-Baxter-Book}: 
\begin{equation}
\label{eq:weightsdef}
\frac{c d}{a b} = \frac{1 - \Gamma}{1 + \Gamma}~;~ 
\frac{a^2 + b^2 - c^2 - d^2}{2 (a b + c d)} = \Delta
\end{equation}
For constant $\Gamma$ and $\Delta$, there exists a one-parameter family of Boltzmann weights ($w$) which satisfy the star–triangle relations and, hence, the eight-vertex model has a one-parameter family of commuting transfer matrices~\cite{2016-Baxter-Book}. This allows one to parameterize the Boltzmann weights explicitly in terms of {\it spectral variable} ($u$):
\begin{equation}
\begin{array}{ll}{a=\operatorname{snh}(\lambda-u)} 
& {b=\operatorname{snh} u}\\ 
{c= k \operatorname{snh} \lambda,} & {d=k \operatorname{snh}(\lambda-u)}\end{array}
\label{elliptic}
\end{equation}
where, $k$ is the elliptic modulus, and $\operatorname{snh}$ is the hyperbolic analogue of $\operatorname{sn}$ and is given by $\operatorname{snh} \ u = -i\ \operatorname{sn}(iu)$.  It has been shown that the transfer matrix of the eight-vertex model commutes with the ${\rm XYZ}$ Hamiltonian~\cite{1970-Sutherland-JMP}:
\begin{equation}
{H}_{\rm XYZ} = -\frac{1}{2} \sum_{j = 1}^N J_{\sigma} \sigma^{\sigma}_j\sigma^{\sigma}_{j+1} 
\quad \mbox{where}  \quad \sigma = x, y, z \, .
\label{XYZ}
\end{equation}
The coupling constants are related to the weights by the relation: 
%
 $J_x:J_y:J_z 
 = 1: \Gamma : \Delta $. 
%
The spins $\sigma_n$'s are related to the vertex weights by the vertex operator ($V_n$):
{\small
$$
 V_n= \frac{1}{2} \left[a+c+ [a-c] \sigma_n^z\sigma_{n+1}^z + 
 [b+d]  \sigma_n^x\sigma_{n+1}^x+ \sigma_n^y\sigma_{n+1}^y\right] 
 $$
}
where, 
\begin{eqnarray}
 a &=& \exp{[K_1+K_2+K'']}, \nonumber \\
 b &=& \exp{[-K_1+K_2+K'']}, \nonumber\\
 c &=& \exp{[-K_1+K_2-K'']}, \nonumber\\
 d &=& \exp{[K_1-K_2-K'']}.   \label{BW}
\end{eqnarray}

\begin{figure}[!htb]
    \centering
    \includegraphics[width=0.75\linewidth]{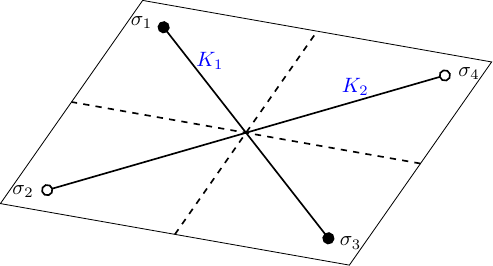}
\caption{A single face of the 8-Vertex Baxter model in the spin formulation. $K_1$ and $K_2$ are the horizontal and vertical spin-spin couplings of the two Ising sublattices.}
\label{fig:Braxtermodel}
\end{figure}

The Hamiltonian of the Baxter model is:
 \begin{equation}
 H_{\rm Baxter}= \sum_{\sigma} \left[ K_1 \, \sigma_1\sigma_3+ K_2 \, \sigma_2\sigma_4 + K'' \, \sigma_1\sigma_2\sigma_3\sigma_4 \right]  \label{H_Baxter}
 \end{equation} 
where $\sigma_1, \sigma_3$  ($\sigma_2, \sigma_4$) are the lattice spins in one (second) Ising sublattice with coupling constant $K_1$ ($K_2$)  as shown in \ref{fig:Braxtermodel}, and $K''$ refers to the four-spin coupling that connects the two sublattices.   

It is possible to obtain the statistical weight (partition function) in the row-to-row transfer matrix and corner transfer matrix formulations~\cite{2016-Baxter-Book}. In our case, it is 
 convenient to use the corner transfer matrix (CTM) formulation as it directly connects the algebra with the conserved charges.  

CTM can be expressed conveniently in terms of Fermions described by the following $XYZ$ spin-chain Hamiltonian~\cite{1986-Thacker-Physica,1987-Itoyama.Thacker-PRL,1989-Itoyama.Thacker-NPB,2016-Baxter-Book}. In this formulation, the partition function is evaluated by choosing a spin in the middle of the lattice and fixing the spins along the vertical and horizontal axes while summing over all spins in the interior of each quadrant~\cite{2016-Baxter-Book}. The final sum over the spins along the axes may be interpreted as the trace of a product of four corner transfer matrices. 

Thus we see that the eight vertex Baxter model is closely related to the $XYZ$ spin chain. Now we can map it to a theory of Fermions by introducing the Corner Transfer Matrix(CTM). 

%
%
%
%

 The CTM operator can be viewed physically as connecting semi-infinite rows of arrows with a semi-infinite column of arrows of one quadrant of the lattice. In the thermodynamic limit, the following relation holds 
\cite{2016-Baxter-Book}:
\begin{equation}
 \mathcal{A}(u)= \exp\left[-\frac{\pi u}{2K} \, \mathcal{L}_0\right] \, , \label{Full A}
\end{equation}
where $K$ is a complete elliptic integral associated with modulus $k$ and 
\begin{equation}
\mathcal{L}_0 = \frac{2K}{\pi}\sum_{j = -\infty}^{\infty}j\mathcal{H}_{XYZ}(j,j+1). \label{L_0}
\end{equation}
$\mathcal{H}_{XYZ}$ corresponds to the Hamiltonian of a spin chain with nearest neighbour (\ref{XYZ})
To keep the calculations transparent, we set $\Gamma = 0$, i. e., $cd = a b$ in \eqref{eq:weightsdef}. This corresponds to the condition $J_z  = 0$ in the Hamiltonian (\ref{XYZ}) which is the well-known XY model~\cite{2016-Baxter-Book}. ${\cal L}_0$ in Eq. (\ref{L_0}) is diagonalized by the operators:
\begin{equation}
{\Psi(l)=N_{l} \int d \alpha \, e^{-\imath \alpha l \pi / 2 K} \chi(\alpha)} 
\label{eq:Psidef}
\end{equation} 
where $N_{l}$ is the normalization constant, and
%
${\chi(\alpha)=\operatorname{sn} \alpha \, a^{\alpha}(\alpha)+\imath \sqrt{k} \operatorname{cn} \alpha \, a^{y}(\alpha)} \, .$
%
The integration over $\alpha$ is over one complete real period of the 
elliptic functions from $-2 K + \imath K'/2$ to $2 K + \imath K'/2$. 
Itoyama and Thacker showed that ${\cal L}_0$ could be expressed as 
\begin{equation}
\mathcal{L}_0 = \sum\limits_l l : \bar{\Psi}(l)\Psi(l) : + \,  h \, ,
\end{equation}
where, $h$ is a constant. ${\cal L}_0$ is embedded into a Virasoro algebra as a central element~\cite{1986-Thacker-Physica,1987-Itoyama.Thacker-PRL,1989-Itoyama.Thacker-NPB}.   The normal ordering is defined by the relations,
$\Psi(l)|h\rangle = 0 (\forall \ l\geq 1),  
\overline{\Psi}(l)|h\rangle = 0, (\forall \ l\leq 1) \, .$ 
%
Other Virasoro operators ${\cal L}_n$ 
can be constructed from these momentum space operators~\cite{1989-Itoyama.Thacker-NPB}.
From \eqref{eq:Psidef}, it follows that,
\begin{equation}
[\mathcal{L}_n,\mathcal{L}_m] = (n-m)\mathcal{L}_{n+m} + \frac{1}{12}c(n^3-n)\delta_{n+m,0}.  \label{L_Virasaro}
\end{equation}
As noted in Ref. \cite{1989-Itoyama.Thacker-NPB}, the physical Hilbert space built from the state $|h\rangle$, forms the highest weight representation of the Virasoro algebra. Since the eigenvalues of $\mathcal{L}_0$ are doubled due to the zero modes of the operator $\Psi(0)$ and $\overline{\Psi}(0)$, the highest weight vector forms a two-dimensional representation under parity conjugation. At the critical point, the central charge $c= 1$ and $h=\frac{1}{8}$.

Using the following classical generators ($l_n^{Diff}$),
\begin{equation}
l_n^{Diff}= -\frac{1}{2}\zeta^{n+1}\frac{d}{d\zeta}- \frac{1}{2}\frac{d}{d\zeta}\zeta^{n+1}. \label{Diffeo_l_n_Defn}
\end{equation}
we can obtain other Virasoro algebra, different from the one described above.  The difference is that $l_n^{Diff}$ are generators of diffeo-transformation of the spectral rapidity parameter or the Euclidean boost parameter ($\alpha$)~\cite{1985-Gervais-PLB,1989-Itoyama.Thacker-NPB}. We can then construct the corresponding Virasoro algebra by defining the following $\mathcal{L}_n^{diff}$:
\begin{eqnarray}
\mathcal{L}_n^{diff} &=& :\int_{-K}^{3K}\frac{d\beta}{2\pi}B(\beta+2K-\imath K')l_nB(\beta):+ h\delta_{n,0} \nonumber\\
&=& \sum\limits_l(l+\frac{n-1}{2}):\bar{\psi}(l)\psi(l+n):+ h \delta_{n,0}. \label{L_N_Diffeo}
\end{eqnarray} 
This demonstrates that the eight-vertex model possesses the Virasoro algebra given by \eqref{L_N_Diffeo}, which holds the key to incorporating near-horizon $S1$ diffeo-symmetry in the model of the horizon-fluid. The 2-D Euclideanized space-time $(\tau,q)$ can be identified with the 2-D  Euclidean space $(x_1,x_2)$ on which the horizon-fluid resides.  The rapidity or the boost parameter in a Euclideanized space-time {corresponds to an} angle of rotation. A closer look reveals that in our case, the rapidity is the azimuthal angle (See  \eqref{fig:Stereographic}). Thus, $l_n^{Diff}$ diffeomorphism algebra of the spectral rapidity corresponds to the $l_n^{Diff}$ diffeomorphism symmetry on the black-hole horizon.  Thus we see that the microscopic modelling of the horizon-fluid with a mass gap incorporates the $l_n^{Diff}$ diffeomorphism symmetry on the black-hole horizon.

\section{Continuum limit and modelling horizon-fluid}

Long-range effects dominate the critical properties of this model; hence, a continuum approximation will suffice for our purpose. The eight-vertex model's continuum limit is a theory of massive Dirac Fermions $(\Psi_1, \Psi_2)$~\cite{1989-Itoyama.Thacker-NPB} 
{
\begin{equation}
\mathcal{S}_{\rm Dirac} =  \int d\tau dq \, \left[\frac{1}{2}i\overline{\Psi}_1\Big(\overleftrightarrow{\frac{\partial}{\partial \tau}}+\overleftrightarrow{\frac{\partial}{\partial q}}\Big)\Psi_1 
+ \frac{1}{2}i\overline{\Psi}_2\Big(\overleftrightarrow{\frac{\partial}{\partial \tau}}-\overleftrightarrow{\frac{\partial}{\partial q}}\Big)\Psi_2 - m(\bar{\Psi}_1\Psi_2+\bar{\Psi}_2\Psi_1) \right]. \label{LagrangianD}
\end{equation}
}
The above action possesses an infinite sequence of conserved densities.
Physically, this implies that besides the total angular momentum, the entire momentum distribution is conserved~\cite{1989-Itoyama.Thacker-NPB}. We can rewrite these operators as integrals of local densities in coordinate space:
\begin{equation}
{\cal L}_{n}=\int dq \,  J_{0}^{(n)}(q) \, , 
\end{equation}
where $J_0$ is the zeroth component of a conserved current. $J^{(-1)}_0$ is the Hamiltonian plus the momentum operator, and $J_0^{(0)}$ (at $\tau = 0$) is the first moment of the Dirac Hamiltonian~\cite{1989-Itoyama.Thacker-NPB}. Integrability ensures the operators are related to the infinite sequence of conserved charges, with one for each new ${\cal L}_n$~\cite{1989-Itoyama.Thacker-NPB}. 

The action \eqref{LagrangianD} is useful to identify the infinite number of conserved charges. However, for our purpose here, starting from the action \eqref{LagrangianD} it is cumbersome to look at the hydrodynamic or long-wavelength properties. Hence, to derive $\zeta$, we will not make direct use of the action \eqref{LagrangianD}. Instead, to make contact with the macroscopic level, i.e., Damour's fluid description, we turn to the fact that the Free-energy density of the Baxter model is the same as that of a classical Ising model in 2-dimensions near the critical point~\cite{2016-Baxter-Book}. On the other hand, the theory of the 2-dimensional Ising model can be described by a theory of a free massive scalar field $\varphi$ in a two-dimensional Euclidean spacetime~\cite{1987-Polyakov-Book}. The key is to identify the microscopic field with the phenomenological order parameter. 

The task of identifying the microscopic field with the phenomenological
order parameter is non-trivial. For the transport theory, we need to describe the dynamics of the black-hole horizon, starting from a $(2+1)-$dimensional Hamiltonian. However, the microscopic model we have constructed is characterized by the {\it 2-D Euclidean space-time Hamiltonian}.  {This means that the microscopic model suffices to describe the quantum states for a system in thermal equilibrium with the desired accuracy but is inadequate if we consider how the system evolves with time. In other words, such Hamiltonians do not contain information about the dynamics of the black-hole horizon. 

There are two ways to tackle this issue. Both approaches keep the thermal states the same but provide different inputs about the dynamics of the system. One way is to assume that a phenomenological Langevin equation gives the dynamics. From this, one can work out the transport theory and compute the correct coefficient of Bulk Viscosity for the Horizon-fluid. The other approach is to construct a $(2+1)-$dimensional Hamiltonian with the same symmetries as the mass gap possessed by the 2-D Euclidean space-time Hamiltonian. For a general process, this is a challenging problem. Fortunately, this difficulty is bypassed for the processes describing bulk viscosity of the fluid using the mean-field theory description. We should emphasize that in this case also, the thermal states remain the same. (The collection of the thermal states of the $(2+1)-$dimensional theory is equivalent to collecting the quantum states of the $1+1$ dimensional Euclideanized theory subject to appropriate boundary conditions.) We shall only discuss the second approach here as a Langevin-type equation can also be derived from the dynamical quantum theory under suitable approximation.} 

The mean-field description will not capture all the details of the symmetries of the quantum states of the model or can reproduce the correct scaling exponents; however, it works satisfactorily to describe horizon-fluid properties.  
Thus, the 2-D mean-field theory Hamiltonian of the microscopic model  
can be extended to the following $2+1$ dimensional space-time Hamiltonian:
\begin{equation}
H_{\rm eff }(\varphi) =   \int \bigg[\frac{1}{2}\big(\frac{\partial \varphi}{\partial t}\big)^2+ \frac{1}{2}\big(\nabla\varphi\big)^2 +\frac{m_{eff}^2}{2}\varphi^2\bigg] dA, \label{H_Th_Defn}
\end{equation}
where the integral is over the area of the event horizon. The above Hamiltonian satisfies the essential requirement of possessing $Z2$ symmetry. Still, it suffers from the weakness that there is very little guidance otherwise in terms of symmetry requirements. One can view 
the above Hamiltonian as a Mean-Field description of the horizon-fluid near a critical point. 

We are now in a position to identify the above Hamiltonian \eqref{H_Th_Defn} with the horizon-fluid.  As a first step, we write down the Landau-Ginzburg expression for the entropy functional corresponding to the horizon-fluid instead of using the Free Energy functional~\cite{1990-Penrose.Fife-PhysicaD}.  This is because the black-hole (of horizon area $A$) constantly interacts with the environment, leading to energy flow.  Thus, we need a framework in which the energy density and the order parameter appear in the formalism on the same footing. As shown in Ref.~\cite{1990-Penrose.Fife-PhysicaD}, the appropriate thermodynamic potential to consider in such a case, is the entropy functional and not free energy.   The stationary state of the black-hole corresponds to the thermodynamic equilibrium state of the black-hole with maximum entropy. The quasi-stationary state of a black-hole corresponds to the vacuum of the deformed CFT.  

Following \cite{2016-Bhattacharya.Shankaranarayanan-PRD}, we assume that the entropy density of the horizon-fluid is homogeneous. We focus on a macroscopically small but finite element of the black-hole horizon area, denoted as ${\cal A}$. ${\cal A}$ satisfies the condition ${\cal A}/A \ll 1$. The order parameter ($\eta)$ is taken to be: 
\[
\eta= C\sqrt{\cal A} \, . 
\]
We can fix the value of the constant $C$ only by relating it to a macroscopic quantity. The entropy functional (${\cal S}$) of the horizon-fluid of this finite element about ($T, {\cal A}$) is given by~\cite{1990-Penrose.Fife-PhysicaD}:
\begin{equation}
{\cal S} = {\cal S}_0(T, {\cal A}) - a \, \eta^2 - b\, \eta^4, \label{SMFT}
\end{equation}
where $a, b$ are constants. 
Let the value of the entropy functional ${\cal S}$ in the quasi-stationary state by ${\cal S}_{QS}$ and the stationary state by ${\cal S}_{S}$, respectively. Note that ${\cal S}_S$ is the global maximum for entropy functional ${\cal S}$. We assume that the process of going from ${\cal S}_{QS}$ to ${\cal S}_{S}$ is a slow physical process so that we can use equilibrium thermodynamics to describe the quasi-stationary state.  For the microscopic model of the horizon-fluid, this means that the ground state of the deformed CFT evolves slowly into a state of CFT.  The slow evolution implies that the deformed CFT vacuum is likely to possess some of the symmetries of a CFT state. This is reminiscent of the adiabatic evolution of a quantum state.

Rewriting (\ref{SMFT}) in terms of the horizon-area at equilibrium, we get,
\begin{eqnarray}
& & ~~~~S_{\rm max}= \frac{{\cal A}_{\rm max}}{4}= {\cal S}_0 -  a \,  C^2 {\cal A}_{\rm max} - b \, C^4  {\cal A}_{\rm max}^2 \, , \nonumber \\
& & \!\!\!\!\!\!\!\!\!
\mbox{where}~~a= -\frac{1}{4C^2}; \quad {\cal S}_0- b\, C^4 \, {\cal A}^2= 0 \, .
\label{a}
\end{eqnarray}
%
Denote the deviation of any variable from its equilibrium value by a prefix $\delta$.
%
%
%
%
The change in the entropy functional is related to the change in the energy density of the horizon-fluid:
\begin{equation}
\delta {\cal H}(\delta\eta) = -T \, \delta {\cal S}  = \frac{T}{2C^2}\delta\eta^2.
\label{H_eff_eta}
\end{equation}
where we have $k_B = 1$. [Expansion around the maximum implies that the terms proportional to $\delta\eta$ cancel out as the expansion is being done around a maximum value of the entropy function. This leads to the condition $2b\eta^2_{max} = -a$. Substituting this in the terms quadratic in $\delta\eta$, i. e.
$(a+2b\eta_{max}^2)\delta\eta^2)$ leads to the above expression.] 

Using the above expression, we can relate the order parameter ($\eta$) with the scalar field  ($\varphi$) and the change in the energy density of the horizon fluid ($\delta {\cal H}(\delta\eta)$) with 
$H_{\rm eff}(\varphi)$ \eqref{H_Th_Defn}, i .e.
\begin{equation}
\delta H = \int dA \, \delta {\cal H} = \delta H_{\rm eff} \, .
\label{eq:HF-Heff}
\end{equation}
For the homogeneous 
process, the spatial gradient terms in the Hamiltonian \eqref{H_Th_Defn} 
can be ignored. Also, the field is taken to vary slowly so the kinetic term
in the Hamiltonian \eqref{H_Th_Defn}  can be ignored. Using the {universality near critical point}, we have~\cite{2016-Baxter-Book,1989-Itoyama.Thacker-NPB}: 
\begin{equation}
\langle\varphi_*\rangle = \frac{\delta\eta}{\sqrt{\cal A}}, 
\label{EtaPhiR}
\end{equation}
where, $\langle\varphi_*\rangle$ is the average value of the field $\varphi_*$ in the area ${\cal A}$. As seen from the above expression, $\varphi_*$ is dimensionless, and is related to $\varphi$ as follows: $\varphi = \sqrt{\Upsilon}\varphi_*$ where $\Upsilon$ has the dimensions of energy. 
The change in the energy of the horizon fluid is given by:
\begin{equation}
\delta H =  \frac{\Upsilon}{2} \, \int dA \, m_{*}^2 \, \varphi_*^2 \, , \quad 
\mbox{where}  \quad m_{*}= \frac{1}{C} \quad \mbox{and} \quad
\Upsilon = T
\label{m_eff}
\end{equation}
Thus, we have made the connection between the microscopic model with 
black-hole thermodynamics. {$m_*$ has the dimensions of energy, i. e.,  $m_* = 1/({l_P^2 \, C})$, $l_P$ being the Planck length. $C$ is dimensionless.  Note that we have set $l_P = 1$.} In the next section, we evaluate the correlations for this process (bulk viscosity). The thermal correlations considered here 
remain the same for Hamiltonian corresponding to action \eqref{LagrangianD} 
and $H_{\rm eff}$. 

\section{Bulk viscosity from the Microscopic model}

We use the above mean-field theory description to calculate the viscosity coefficient of the horizon-fluid from the correlations of the energy-momentum tensor of the scalar field given by the Hamiltonian \eqref{H_Th_Defn}. Jeon has developed such a description~\cite{JeonFluid}. First, however, we need to make suitable changes to apply to the horizon-fluid. 

According to Jeon, the coefficient of bulk-viscosity ($\zeta$) of a viscous-fluid with stress-tensor is~\cite{JeonFluid}:
\begin{equation}
\zeta = \frac{\beta}{2}\lim_{\omega\rightarrow 0}\lim_{{\mathbf q}\rightarrow 0}\sigma_{\bar{P}\bar{P}}, \label{zetasigma}
\end{equation}
where, $\sigma_{\bar{P}\bar{P}}$ is given by 
\begin{equation}
\sigma_{\bar{P}\bar{P}}(\omega,{\mathbf q})= \frac{1}{2\pi {\cal A}} \int d^2{\mathbf x}\int_{-\infty}^{\infty}dt e^{-i{\mathbf q}.{\mathbf x}+i\omega t}\langle\bar{P}(t,{\mathbf x})\bar{P}(0)\rangle, \label{sigma_P}
\end{equation}
where ${\cal A}$ is the area normalization of the spatial part,
\begin{equation}
\bar{P}(t,{\mathbf x})= P(t,{\mathbf x}) - v_S^2\rho(t,{\mathbf x})= \frac{1}{2}T_i^i(x^\mu)- v_S^2T_{00}(x^\mu). \label{PbarDef}
\end{equation}
and $v_S$ is the speed of sound of the field.

To apply this formalism to horizon-fluid, we need to make suitable changes in the formulation. As mentioned earlier,
the stress-tensor of the horizon-fluid vanishes as the infalling matter-energy reaches the horizon and the horizon becomes quasi-stationary~\cite{1982-Damour-Proc,1986-Price.Thorne-PRD,1986-Thorne.etal-Membrane}. In other words, when the matter reaches an equilibrium at a given temperature, the stress-tensor of the horizon-fluid is zero.  Thus, the field-theoretic description of the horizon-fluid corresponds to the deviation of the energy-momentum tensor of the field ($T_{\mu\nu}^H$) 
from its thermal average at the thermal state ($\langle T_{\mu\nu}^H\rangle_{\rm equilibrium}$). In the case of normal-fluid, we use the stress-tensor (cf. \ref{sigma_P}). For the horizon-fluid on the other hand, the key quantity is the deviation of the energy-momentum tensor of the field, i. e.,
\[
\delta T_{\mu\nu}^H= T_{\mu\nu}^H- \langle T_{\mu\nu}^H\rangle_{\rm equilibrium} \, .
\]
Physically, this corresponds to the state when the expectation value of the stress-energy tensor of the perturbed CFT on the horizon is the thermal average.  
Thus, for the horizon-fluid, Eqs.~\eqref{sigma_P} and \eqref{PbarDef} become:
\begin{eqnarray}
\label{delta Pbar}
\delta\bar{P(t, {\mathbf x})} &=&  \frac{1}{2}\delta T_i^i(x^\mu)- \delta T_{00}(x^\mu) \\ 
\sigma_{\delta\bar{P},\delta\bar{P}}(\omega,{\mathbf q}) &=& 
 \frac{1}{2\pi {\cal A}} \int d^2{\mathbf x}\int_{-\infty}^{\infty}dt 
 e^{-i{\mathbf q} \cdot {\mathbf x}+i\omega t}\langle\delta\bar{P}(t,{\mathbf x})\delta\bar{P}(0)\rangle_0 . \label{sigmadelta}
\end{eqnarray}
For the horizon-fluid, $v_S$ is equal to the speed of light $(c = 1$). For the Hamiltonian \eqref{H_Th_Defn}, we have:
\begin{equation}
\bar{P}= -\frac{1}{2}(\nabla\phi)^2- m_{\rm eff}^2\phi^2. \label{PBar}
\end{equation}
We can ignore contributions from the $(\nabla\phi)$ term for the homogeneous perturbations responsible for the bulk viscosity. We can determine $\delta T_{\mu\nu}$ by systematically tracking the deviation of the field $\phi$ from its average value at equilibrium state, i .e., 
\[
\phi= \langle\phi\rangle_0 + \delta\phi
\]
where $\langle\rangle_0 \equiv \phi_0$ denotes the ensemble average of the density matrix.

For a field at a temperature $T$, the density matrix is the thermal density matrix at $T$. In the hydrodynamic limit, $
\phi_0$ is given by:
\begin{equation}
\phi_0^2 = \frac{\sum e^{-\beta H} \phi^2}{\sum e^{-\beta H}} = \frac{\int \mathcal{D}\phi \phi^2 e^{-\beta \int \frac{m_{\rm eff}^2}{2}\phi^2 d^2x}}{\int \mathcal{D}\phi e^{-\beta \int \frac{m_{\rm eff}^2}{2}\phi^2 d^2x}}=\frac{1}{2m_*^2}, \label{phi_0}
\end{equation}
where star denotes that rescaled variables [see the discussion below Eq. \eqref{HTotal}]. Writing $\phi=\phi_0+\delta\phi$, \eqref{PBar} can be written as
\begin{equation}
\delta\bar{P}= -2m_*^2\phi_0\delta\phi= -2\frac{m_*}{\sqrt{2}}\delta\phi= \sqrt{2}m_*\delta\phi= \frac{\sqrt{2}}{C}\delta\phi. \label{DeltaP2}
\end{equation}
The Coefficient of Bulk Viscosity is given by,
\begin{equation}
\zeta = 2\pi \operatorname{Im} \Big[\frac{1}{4 \pi {\cal A}}\lim_{{\mathbf q}\rightarrow 0}\lim_{\omega\rightarrow m_*}\int d^3x e^{-i{\mathbf q}.{\mathbf x}+i\omega t}\langle\big[\delta\bar{P}(t,\mathbf{x}), \delta\bar{P}(0)\big]\rangle\theta(-t)\Big]. \label{ZetaA01}
\end{equation}
Substituting \eqref{DeltaP2} in \eqref{ZetaA01}, and keeping in mind that the perturbations are homogeneous, one gets, 
\begin{equation}
\zeta = \operatorname{Im} \Big[\lim_{\omega\rightarrow m_*}\frac{1}{C^2}\int_{-\infty}^{\infty} dt e^{i\omega t}\langle\big[\delta\hat{\phi}(t),\delta\hat{\phi}(0)\big]\rangle\theta(-t)\Big]. \label{ZetaF}
\end{equation}
$\langle\big[\delta\hat{\phi}(t,\mathbf{x}),\delta\hat{\phi}(0)\big]\rangle\theta(-t)$ corresponds to the advanced Green's function that appears because of the teleological nature of the event horizon [for detailed discussion, see~\cite{1986-Price.Thorne-PRD,1986-Thorne.etal-Membrane,2016-Bhattacharya.Shankaranarayanan-PRD}], and
\begin{equation}
\big[\delta\hat{\phi}(t),\delta\hat{\phi}(0)\big]= \big[\hat{\phi}(t),\hat{\phi}(0)\big]
\label{deltaphiOp}
\end{equation}
To obtain the above commutation relation of the field operator, we write the Hamiltonian \eqref{H_Th_Defn} as
\begin{eqnarray}
 H_{\rm HF} = 
\frac{T}{2} \int \bigg[\big(\frac{\partial \varphi_*}{\partial t}\big)^2+ \big(\nabla\varphi_*\big)^2 + m_*^2\varphi_*^2
 \bigg]  
dA \, . \label{HTotal}
\end{eqnarray}
where $\phi= \sqrt{T}\phi_*$, and $\phi_*$ is the rescaled (dimensionless) effective scalar field and $T$ is taken to be a constant for convenience. Rewriting $\hat{\varphi}_*$ (also $\hat{\varphi}$) in terms of the creation and annihilation operators, i. e.,
\begin{equation}
\hat{\varphi_*}(t,\mathbf{x})= \sum_{\mathbf{k_*}}[\hat{a}_{\mathbf{k_*}}u_{\mathbf{k_*}}(t,\mathbf{x})+ \hat{a}_{\mathbf{k_*}}^{\dagger}u_{\mathbf{k_*}}^*(t,\mathbf{x})],
\label{eq:modeexpansion}
\end{equation}
where, 
\begin{equation}
u_{\mathbf{k_*}}(t,\mathbf{x}) = \frac{1}{\sqrt{2\pi}}\frac{1}{\sqrt{2\omega_*}} e^{i(\mathbf{k_*}.\mathbf{x}-\omega_* t)}
\end{equation}
and $\mathbf{k_*}$, $\omega_*$, $\mathbf{x}$ and $t$ are dimensionless variables. Expressing the dimensionless part of the above Hamiltonian \eqref{HTotal} in the frequency domain, we have:
\begin{equation}
H_{\rm eff}(\varphi)= T \sum_{\mathbf{k_*}}(\hat{a}_{\mathbf{k_*}}^{\dagger}\hat{a}_{\mathbf{k_*}}+\frac{1}{2})\omega_* , 
\end{equation}
where the dispersion relation is given by:
\begin{equation}
\omega_*^2 = \mathbf{k_*}^2+ \frac{1}{C^2} \quad {\rm and} 
\quad C=\frac{1}{m_*} \, .
\label{Dispersion}
\end{equation}
In the hydrodynamic limit ($\mathbf{k_*}\rightarrow 0$), the above expression reduces to $\omega_{*H} = {1}/{C}$. Substituting Eq.~\eqref{deltaphiOp} in Eq. \eqref{ZetaF} and after a little algebraic manipulation in terms of dimensionless variables, we have~\cite{1966-Kubo-RPP}: 
\begin{equation}
\zeta[\omega_{*H}] = \operatorname{Im}\left[ - \frac{i}{4C^2} 
\frac{T\omega_{*H}}{E_\beta(\omega_{*H})}\int_{-\infty}^{\infty}\langle
\hat{\varphi}(0)\hat{\varphi}(t) + \hat{\varphi}(t)\hat{\varphi}(0)
\rangle e^{-i\omega_{*H} t}\theta(-t) dt \right], 
\label{Zeta2a}
\end{equation}
where 
$E_\beta(\omega_{*H})$ is the average energy of excitation in the mode with frequency $\omega_{*H}$ at temperature $T$. The above integral exhibits a pole at $\omega_*' = \omega_{*H}$, which is the well-known pole at the hydrodynamical limit. Using the fact that the process corresponding to bulk viscosity is  homogeneous, we have
\begin{equation}
\langle\hat{\varphi}^2_{\omega_H}(0)\rangle =
\frac{1}{(2\pi)^2}\frac{1}{4m_*^2}= \frac{1}{(2\pi)^2} \frac{C^2}{4} \label{eqphio2}
\end{equation}
In the hydrodynamic limit, we get: 
\begin{equation}
\label{Zeta6}
\zeta[\omega_{*H}] = -\frac{1}{128\pi^2}\frac{T\omega_{*H}}{E_\beta(\omega_{*H})}= -\frac{1}{128\pi^2} 2 \tanh{\frac{1}{2C}} \, .
\end{equation}
Demanding that the above expression matches with the expression 
derived by Damour~\cite{1982-Damour-Proc}:
\[
\zeta= - \frac{1}{16\pi} \, ,
\]
we have 
\begin{equation}
\tanh{\frac{1}{2C}}= \frac{1}{4\pi} .
\end{equation}
Solving this equation numerically, we get,$C= 6.2696$ and hence, $m_*= 0.1595$. The negative sign of a transport coefficient arises due to $\theta(-t)$ in the response function of the black-hole horizon~\cite{2016-Bhattacharya.Shankaranarayanan-PRD}. 

\section{Discussion}

The minimal microscopic model for the horizon-fluid discussed here incorporates three simple ideas. First, the model contains an infinite-dimensional symmetry algebra, namely the Virasaro algebra, corresponding to black-holes' near-horizon $S1$ diffeo-symmetry.  Second, the model describes a perturbed CFT that possesses a representation of the Virasoro algebra. The deformation is characterized by a mass term and is supposed to address the perturbed nature of the non-stationary black-hole horizon. Finally, the mass gap provides a simple derivation of $\zeta$ and area quantization, thus demonstrating that in the long-wavelength limit, it can connect with the known semiclassical Physics of black-holes.  Historically, Bekenstein was the first to derive the quantization of area using adiabatic quantization. We obtained quantization of entropy by slowly evolving the black-hole horizon. Adiabatic quantization of a system is applicable when some of the parameters characterizing the system evolve slowly.

In this talk, only the homogeneous perturbations of the stationary black-hole have been considered. We need to construct a microscopic model, including general perturbations, which can describe the horizon-fluid. This complete theory would, of course, be richer than the toy model we have put forward. So the Baxter model serves only as an illustration in this sense. Nonetheless, an improved model should also share the first three critical features of the current model mentioned above.  Thus, we expect a more comprehensive microscopic future model of the horizon-fluid to describe a system close to the critical point. It should represent the near-horizon symmetry algebra and a mass gap with a fixed, known value.

Our results also lead to an interesting suggestion. The twin requirements of the theory incorporating near-horizon symmetries($S1$ diffeomorphism) and possessing length scales due to external perturbations can be naturally satisfied if the theory on the non-stationary black-hole horizon is a deformed CFT. Similar types of arguments also lead to the expectation that the theory on the stationary black-hole horizon is a CFT~\cite{1999-Carlip-PRL}. Since the non-stationary black-hole eventually becomes stationary, it is reasonable to assume that the deformed CFT flows into a CFT as the black-hole horizon dynamically evolves. Note that the above argument does not depend on any specific model. This leads us to conjecture that \emph{the low energy theory on the black-hole horizon flows towards a critical point as the perturbed black-hole ultimately becomes stationary.}  We think that this conjecture may be a good starting point to construct quantum theory models of the black-hole horizon.

\bibliographystyle{ws-procs961x669}


\begin{thebibliography}{10}

\bibitem{2001-Wald-LRR}
Robert~M. Wald.
\newblock The thermodynamics of black holes.
\newblock {\em Liv. Rev. Rela.}, 4:6, 2001.

\bibitem{2003-Jacobson.Parentani-FP}
Ted Jacobson and Renaud Parentani.
\newblock Horizon entropy.
\newblock {\em Found. Phys.}, 33(2):323--348, 2003.

\bibitem{2008-Carlip-Lec}
Steven Carlip.
\newblock {Black Hole Thermodynamics and Statistical Mechanics}.
\newblock {\em Lect. Notes Phys.}, 769:89--123, 2009.

\bibitem{1973-Bardeen.etal-CMP}
James~M. Bardeen, B.~Carter, and S.~W. Hawking.
\newblock The four laws of black hole mechanics.
\newblock {\em Commun. Math. Phys.}, 31:161--170, 1973.

\bibitem{1995-Jacobson-PRL}
Ted Jacobson.
\newblock {Thermodynamics of space-time: The Einstein equation of state}.
\newblock {\em Phys. Rev. Lett.}, 75:1260--1263, 1995.

\bibitem{2010-Padmanabhan-RPP}
T.~{Padmanabhan}.
\newblock {Thermodynamical aspects of gravity: new insights}.
\newblock {\em Rep. Prog. Phys.}, 73(4):046901, April 2010.

\bibitem{2015-Padmanabhan-Entropy}
T.~{Padmanabhan}.
\newblock {Distribution Function of the Atoms of Spacetime and the Nature of
  Gravity}.
\newblock {\em Entropy}, 17:7420--7452, October 2015.

\bibitem{1975-Hawking-CMP}
S.~W. Hawking.
\newblock Particle creation by black holes.
\newblock {\em Commun. Math. Phys.}, 43:199--220, 1975.

\bibitem{2007-Das.Shankaranarayanan-CQG}
Saurya Das and S.~Shankaranarayanan.
\newblock Where are the black hole entropy degrees of freedom?
\newblock {\em Class. Quant. Grav.}, 24:5299--5306, 2007.

\bibitem{2019-Romatschke.Romatschke-Book}
Paul Romatschke and Ulrike Romatschke.
\newblock {\em {Relativistic Fluid Dynamics In and Out of Equilibrium}}.
\newblock Cambridge Monographs on Mathematical Physics. Cambridge University
  Press, 2019.

\bibitem{1982-Damour-Proc}
T.~{Damour}.
\newblock {Surface Effects in Black-Hole Physics}.
\newblock In R.~{Ruffini}, editor, {\em Marcel Grossmann Meeting: General
  Relativity}, 1982.

\bibitem{1986-Price.Thorne-PRD}
R.~H. Price and K.~S. Thorne.
\newblock {Membrane viewpoint on black holes: Properties and evolution of the
  stretched horizon}.
\newblock {\em Phys. Rev. D.}, 33:915--941, February 1986.

\bibitem{1986-Thorne.etal-Membrane}
Kip~S Thorne, Richard~H Price, and Douglas~A Macdonald.
\newblock {\em The Membrane Paradigm}, volume~19.
\newblock Yale University Press, New Haven, 1986.

\bibitem{1998-Kaul.Majumdar-PLB}
Romesh~K. Kaul and Parthasarathi Majumdar.
\newblock {Quantum black hole entropy}.
\newblock {\em Phys. Lett.}, B439:267--270, 1998.

\bibitem{2011-Kaul.Majumdar-PRD}
Romesh~K. Kaul and Parthasarathi Majumdar.
\newblock {Schwarzschild horizon dynamics and SU(2) Chern-Simons theory}.
\newblock {\em Phys. Rev.}, D83:024038, 2011.

\bibitem{2001-Koga-PRD}
Jun-ichirou Koga.
\newblock {Asymptotic symmetries on Killing horizons}.
\newblock {\em Phys. Rev. D}, D64:124012, 2001.

\bibitem{2001-Hotta.etal-CQG}
M.~Hotta, K.~Sasaki, and T.~Sasaki.
\newblock {Diffeomorphism on horizon as an asymptotic isometry of Schwarzschild
  black hole}.
\newblock {\em Class. Quant. Grav.}, 18:1823--1834, 2001.

\bibitem{2002-Hotta-PRD}
M.~Hotta.
\newblock {Holographic charge excitations on horizontal boundary}.
\newblock {\em Phys. Rev.}, D66:124021, 2002.

\bibitem{1999-Carlip-PRL}
S.~Carlip.
\newblock Black hole entropy from conformal field theory in any dimension.
\newblock {\em Phys. Rev. Lett.}, 82:2828, 1999.

\bibitem{2011-Carlip-Entropy}
Steven Carlip.
\newblock {Effective Conformal Descriptions of Black Hole Entropy}.
\newblock {\em Entropy}, 13:1355--1379, 2011.

\bibitem{2016-Bhattacharya.Shankaranarayanan-PRD}
Swastik Bhattacharya and S.~Shankaranarayanan.
\newblock {Fluctuations in horizon-fluid lead to negative bulk viscosity}.
\newblock {\em Phys. Rev. D}, D93(6):064030, 2016.

\bibitem{2017-Cropp.etal-PRD}
Bethan Cropp, Swastik Bhattacharya, and S.~Shankaranarayanan.
\newblock Hints of quantum gravity from the horizon fluid.
\newblock {\em Phys. Rev. D.}, D95(2):024006, 2017.

\bibitem{1992-Chandrasekhar-BHBook}
S.~Chandrasekhar.
\newblock {\em The mathematical theory of black holes}.
\newblock Oxford Univ. Press, Oxford, 1992.

\bibitem{2010-Barnich.Troessaert-PRL}
Glenn Barnich and Cedric Troessaert.
\newblock {Symmetries of asymptotically flat 4 dimensional spacetimes at null
  infinity revisited}.
\newblock {\em Phys. Rev. Lett.}, 105:111103, 2010.

\bibitem{2016-Donnay.etal-PRL}
Laura Donnay, Gaston Giribet, Hernan~A. Gonzalez, and Miguel Pino.
\newblock {Supertranslations and Superrotations at the Black Hole Horizon}.
\newblock {\em Phys. Rev. Lett.}, 116(9):091101, 2016.

\bibitem{1962-Bondi.etal-PRLSA}
H.~Bondi, M.~G.~J. van~der Burg, and A.~W.~K. Metzner.
\newblock {Gravitational waves in general relativity. 7. Waves from
  axisymmetric isolated systems}.
\newblock {\em Proc. Roy. Soc. Lond.}, A269:21--52, 1962.

\bibitem{1962-Sachs-PRLSA}
R.~K. Sachs.
\newblock {Gravitational waves in general relativity. 8. Waves in
  asymptotically flat space-times}.
\newblock {\em Proc. Roy. Soc. Lond.}, A270:103--126, 1962.

\bibitem{2017-Strominger-Arx}
Andrew Strominger.
\newblock {Lectures on the Infrared Structure of Gravity and Gauge Theory}.
\newblock 2017.

\bibitem{2012-Compere-LRR}
Geoffrey Comp\'ere.
\newblock {The Kerr/CFT correspondence and its extensions}.
\newblock {\em Living Rev. Rel.}, 15:11, 2012.

\bibitem{1987-Zamolodchikov-JETPL}
AB~Zamolodchikov.
\newblock Higher-order integrals of motion in two-dimensional models of the
  field theory with a broken conformal.
\newblock {\em JETP Lett}, 46(4), 1987.

\bibitem{1989-Zamolodchikov-Proc}
Alexander~B Zamolodchikov.
\newblock Integrable field theory from conformal field theory.
\newblock In {\em Integrable Systems and Quantum Field Theory}, pages 641--674.
  Elsevier, 1989.

\bibitem{1971-Baxter-PRL}
Rodney~J Baxter.
\newblock Eight-vertex model in lattice statistics.
\newblock {\em Physical Review Letters}, 26(14):832, 1971.

\bibitem{1973-Baxter.Wu-PRL}
RJ~Baxter and FY~Wu.
\newblock Exact solution of an ising model with three-spin interactions on a
  triangular lattice.
\newblock {\em Physical Review Letters}, 31(21):1294, 1973.

\bibitem{1976-Baxter-JStatPhys}
R.~J. Baxter.
\newblock Corner transfer matrices of the eight-vertex model. i.
  low-temperature expansions and conjectured properties.
\newblock {\em Journal of Statistical Physics}, 15(6):485--503, Dec 1976.

\bibitem{1977-Baxter-JStatPhys}
R.~J. Baxter.
\newblock Corner transfer matrices of the eight-vertex model. ii. the ising
  model case.
\newblock {\em Journal of Statistical Physics}, 17(1):1--14, Jul 1977.

\bibitem{1978-Baxter-JStatPhys}
Rodney~J Baxter.
\newblock Variational approximations for square lattice models in statistical
  mechanics.
\newblock {\em Journal of Statistical Physics}, 19(5):461--478, 1978.

\bibitem{2016-Baxter-Book}
Rodney~J Baxter.
\newblock {\em Exactly solved models in statistical mechanics}.
\newblock Elsevier, 2016.

\bibitem{1971-Kadanoff.Wegner-PRB}
Leo~P Kadanoff and Franz~J Wegner.
\newblock Some critical properties of the eight-vertex model.
\newblock {\em Physical Review B}, 4(11):3989, 1971.

\bibitem{1987-Itoyama.Thacker-PRL}
H.~Itoyama and H.~B. Thacker.
\newblock {Lattice Virasoro Algebra and Corner Transfer Matrices in the Baxter
  Eight Vertex Model}.
\newblock {\em Phys. Rev. Lett.}, 58:1395, 1987.

\bibitem{1986-Thacker-Physica}
HB~Thacker.
\newblock Corner transfer matrices and lorentz invariance on a lattice.
\newblock {\em Physica D: Nonlinear Phenomena}, 18(1-3):348--359, 1986.

\bibitem{1989-Itoyama.Thacker-NPB}
H.~Itoyama and H.~B. Thacker.
\newblock {Integrability and Virasoro Symmetry of the Noncritical Baxter-Ising
  Model}.
\newblock {\em Nucl. Phys.}, B320:541--590, 1989.

\bibitem{JeonFluid}
S.~Jeon,
Phys. Rev. D \textbf{52}, 3591-3642 (1995)
doi:10.1103/PhysRevD.52.3591
[arXiv:hep-ph/9409250 [hep-ph]].

\bibitem{1970-Sutherland-JMP}
Bill Sutherland.
\newblock Two-dimensional hydrogen bonded crystals without the ice rule.
\newblock {\em Journal of Mathematical Physics}, 11(11):3183--3186, 1970.

\bibitem{1985-Gervais-PLB}
J.-L. Gervais.
\newblock Infinite family of polynomial functions of the virasoro generators
  with vanishing poisson brackets.
\newblock {\em Physics Letters B}, 160(4):277 -- 278, 1985.

\bibitem{1987-Polyakov-Book}
A~M. Polyakov.
\newblock {\em Gauge Fields and Strings}.
\newblock Contemporary concepts in physics. Taylor \& Francis, 1987.

\bibitem{1990-Penrose.Fife-PhysicaD}
Oliver Penrose and Paul~C Fife.
\newblock Thermodynamically consistent models of phase-field type for the
  kinetic of phase transitions.
\newblock {\em Physica D}, 43(1):44--62, 1990.

\bibitem{1966-Kubo-RPP}
R.~{Kubo}.
\newblock {The fluctuation-dissipation theorem}.
\newblock {\em Reports on Progress in Physics}, 29:255--284, January 1966.

\bibitem{1957-Kubo}
R.~Kubo,
J. Phys. Soc. Jap. \textbf{12}, 570-586 (1957)
doi:10.1143/JPSJ.12.570

\bibitem{1974-Bekenstein-NCL}
J.~D. Bekenstein.
\newblock The quantum mass spectrum of the kerr black hole.
\newblock {\em Lettere al Nuovo Cimento (1971-1985)}, 11(9):467--470, 1974.

\end{thebibliography}
\end{document}